\documentclass[mathleft]{an}
\usepackage{graphicx}
\usepackage{times}
\usepackage{lscape}
\usepackage{longtable}
\newcommand{\ltsim}{\protect\raisebox{-0.5ex}{$\:\stackrel{\textstyle <}
        {\sim}\:$}}

\begin{document}

\Pagespan{1}{7}
\Yearpublication{}
\Yearsubmission{}
\Month{}   
\Volume{}  
\Issue{} 

\title{Effective temperature vs line-depth ratio for ELODIE spectra.\\
Gravity and rotational velocity effects\,\thanks{ Based on spectral 
data retrieved from the ELODIE Archive at Observatoire de Haute-Provence (OHP).}}

\author{K. Biazzo\inst{1,2}\fnmsep\thanks{Corresponding author: 
\email{katia.biazzo@oact.inaf.it}.\newline}
  \and A. Frasca\inst{1}  \and S. Catalano\inst{1}  \and E. Marilli\inst{1}}

\titlerunning{Effective temperature vs line-depth ratio for ELODIE spectra}
\authorrunning{K. Biazzo et al.}

\institute{
INAF - Catania Astrophysical Observatory, via S. Sofia 78, I-95123 Catania, Italy
\and
Department of Physics and Astronomy, University of Catania, via S. Sofia 78, I-95123 Catania, Italy
}

\received{}
\accepted{}
\publonline{}


\abstract{The dependence on the temperature of photospheric line-depth ratios (LDRs) in the spectral range 
6190--6280 \AA~is investigated by using a sample of 174 ELODIE Archive stellar spectra of luminosity 
class from V to III.  The rotational broadening effect on LDRs is also studied. We provide useful 
calibrations of effective temperature versus LDRs for giant and main sequence stars with 
$3800\ltsim T_{\rm eff}\ltsim 6000$ K and $v\sin i$ in the range 0--30 km s$^{-1}$. We found that, with 
the exception of very few line pairs, LDRs, measured at a spectral resolution as high as 42\,000, 
depend on $v\sin i$ and that, by neglecting the rotational broadening effect, one can mistake the 
$T_{\rm eff}$ determination of $\sim$100 K in the worst cases.}

\maketitle

\section{Introduction}
The stellar effective temperature is one of the most important astrophysical parameters. It is not easy to measure it 
with high accuracy and this situation worsens for stars that have already left the main sequence and that 
are expanding their envelopes and redistributing their angular momentum. In addition, 
surface inhomogeneities, such as spots and faculae can significantly affect the average temperature of 
active stars. 

Several diagnostics, based on color indices and spectral-type classification, are available for temperature measurements, 
but errors greater than 100 Kelvin degrees are often encountered. On the other hand, the analysis 
of line profiles and their dependence upon the stellar effective temperature is a very powerful tool for the 
study of the photospheric temperature, as well as for the investigation of stellar surface structures.

As a matter of fact, it has been demonstrated that the ratio of the depths of two lines having different sensitivity 
to temperature is an excellent diagnostics for measuring small temperature differences between stars or small 
temperature variations of a given star. Although the effective temperature scale can be set to within a few tens of degrees, 
temperature differences can be measured with a precision down to a few Kelvin degrees in the most favorable cases 
(e.g., Gray \& Johanson 1991, Strassmeier \& Schordan 2000, Gray \& Brown 2001, Catalano et al. 2002a, 2002b). This 
allows putting a star sample in a ``relative'' temperature scale or to detect tiny temperature variations in individual stars. 
Starspot temperatures from line-depth ratios (LDRs) variations, indeed, have been determined in the slowly rotating dwarf 
star $\sigma$~Draconis (Gray et al. 1992), in the very young and rapidly rotating star LQ~Hydrae (Strassmeier et al. 1993), 
in some young solar-type stars (Biazzo et al. 2007), and in three single-lined RS~CVn binaries (Frasca et al. 2005). 
However, all these works do not explicitly take into account the effects of stellar rotation on spectral line depths 
and consequently on LDRs.

In the present work, we primarily explore the influence of the rotational velocity on LDRs and provide calibration relations 
between individual LDRs and effective temperature at different rotation velocities for high-resolution spectra 
($R=42\,000$) taken from the ELODIE Archive (Moultaka et al. 2004), taking also into account the effect of gravity. 

The effect of slightly different metallicity for the stars in our sample is also examined. Its influence on 
LDRs is found to be negligible for stars around solar metallicity (within $\pm$0.3).

\section{The star sample}
\subsection{Sample selection}
We selected 174 stellar spectra of giant and main sequence stars (MS) from the on-line ELODIE Archive 
(Moul\-taka et al. 2004). Their spectral types range from F5 to M0 (Hoffleit \& Warren 1991). The criteria 
for the choice of the stellar sample are the following: 
\begin{itemize}
\item low rotation rate, $v\sin i\ltsim$ 5 km s$^{-1}$ (values mainly taken from Bernacca \& Perinotto 1970, 
de Medeiros \& Mayor 1999, Glebocki et al. 2000);
\item good Hipparcos parallaxes ($\pi$) with errors less than 15\% (ESA 1997);
\item accurate $B-V$ color indices (Mermilliod \& Mermilliod 1998) ranging from 0.49 to 1.37 for 
the 102 MS stars and from 0.64 to 1.54 for the 72 giant stars;
\item nearly solar metallicity, [Fe/H]=$0.0\pm$0.3 (Soubiran et al. 1998).
\end{itemize}

\subsection{Effective temperature and luminosity}
\begin{figure}
{\includegraphics[width=8cm]{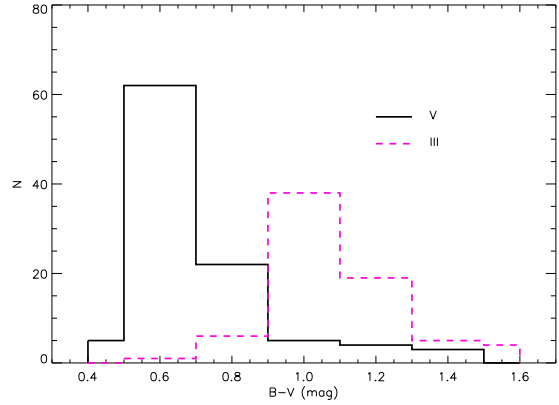}}
\caption{$B-V$ distribution of the giant (III) and MS (V) calibration stars.}
\label{fig:hist_standard_III_V}
\end{figure}

In order to determine the $T_{\rm eff}$ and luminosity of the stars in our sample and to put them onto the HR
diagram, it is necessary to evaluate the interstellar extinction ($A_V$) and reddening ($E_{B-V}$). We have evaluated $A_V$ 
from the star distance, assuming a mean extinction of 1.7 mag/kpc for stars on the galactic plane ($|b| < 5\degr$) and of 0.7 
mag/kpc for stars out of the plane ($|b| > 5\degr$). The reddening was estimated according to the standard law 
$A_{V}= 3.1\,E_{B-V}$ (Savage \& Mathis 1979), with $E_{B-V}$ color excess. However, the color excess of these 
nearby stars is always less than 0\fm03 for MS stars and less than 0\fm09 for giant stars. 
The only two exceptions are HD 176737 and HD 54489 where $E_{B-V}\sim 0.20$ since they have $|b| < 5\degr$ 
and $\pi < 3$ mas. Fig.~\ref{fig:hist_standard_III_V} shows the distributions of the de-reddened color $(B-V)_0$ 
for the MS and giant stars. The de-reddened $V$ magnitude was converted into absolute magnitude 
$M_{V}$ through the parallax and subsequently into bolometric magnitude by means of the bolometric corrections tabulated 
by Flower (1996) as a function of the effective temperature. A solar bolometric magnitude of $M_{\rm bol} = 4.64$ (Cox 2000) 
was used to express the stellar luminosity in solar units. From $(B-V)_0$ values we have deduced the effective 
temperature by means of the following empirical relation given by Gray (2005) and valid for 
0.00$\le (B-V)_0\ltsim$1.5:
\begin{eqnarray}
\label{eq:Gray_cal}
\log T_{\rm eff} & = & 3.981-0.4728(B-V)_0\nonumber \\
		 & + & 0.2434(B-V)_0^2 - 0.062(B-V)_0^3
\end{eqnarray}

\begin{figure}
{\includegraphics[width=8cm]{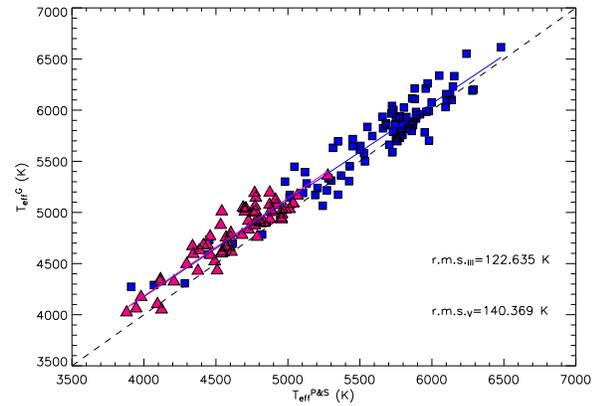}}
\caption{Comparison between the effective temperatures of the giant (triangles) and MS (squares) stars obtained using 
the Gray's calibration (Gray 2005) and those listed by Prugniel \& Soubiran (2001). The two continuous lines represent the 
linear fits to the two star groups, while the dashed line is the bisector.}
\label{fig:Teff_Gray_Prugniel}
\end{figure}

We have verified the consistency of  $T_{\rm eff}$ values derived in this way with those listed by Prugniel \& Soubiran (2001) 
and compiled from the literature (Fig.~\ref{fig:Teff_Gray_Prugniel}). The latter temperatures appear to 
be systematically lower than Gray's temperatures by about 120~K, on average. The root mean 
square (rms) of data compared to the linear fits is around 130~K both for giant and MS 
stars. This comparison allows us to estimate an average temperature uncertainty of about 100--150~K 
for the FGK stars of our sample, as already found in previous studies (see, e.g., Gray 2005, and 
references therein). We find analogous results by plotting the $T_{\rm eff}$ derived from Gray's relation 
with those obtained by means of the $T_{\rm eff}$--$(V-I)$ calibration by Alonso et al. (1996, 1999).
After several tests made on the LDR--$T_{\rm eff}$ calibrations, we have chosen to use 
Gray's effective temperatures for the following analysis, because they give the smallest scatter in the calibrations 
(rms$\simeq$0.13, 0.14, and 0.06 for Prugniel \& Soubiran, Alonso et al., and Gray temperature sets, respectively). 
Anyway, the use of Alonso et al.'s temperature scale, instead of Gray's scale, does not affect significantly 
the LDR--$T_{\rm eff}$ calibrations. 

\begin{figure}
{\includegraphics[width=8cm]{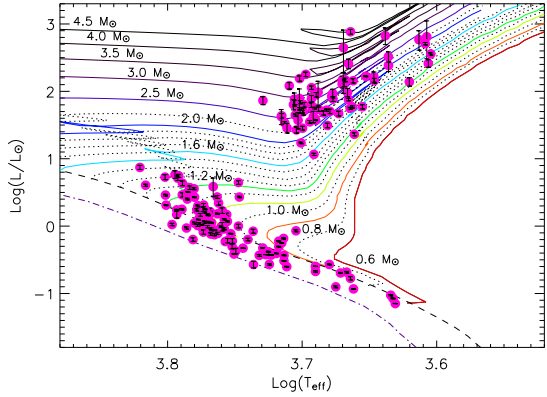}}
\caption{HR diagram of the stars in our sample. The evolutionary tracks for different masses are taken 
from Girardi et al. (2000) and are shown by continuous and dotted lines. The isochrones at an 
age of 6.31$\times$10$^7$ yrs (ZAMS) with $Z=0.019$ ([Fe/H]=0) and $Z=0.01$ ([Fe/H]$\sim-$0.3) are 
also displayed with dashed and dash-dotted lines, respectively.}
\label{fig:hr_diagram_elo_red}
\end{figure}

The position onto the Hertzsprung-Russel (HR) diagram of the ELODIE stars used in this work is shown in 
Fig.~\ref{fig:hr_diagram_elo_red} together with the evolutionary tracks and isochrones calculated by 
Girardi et al. (2000). 

\section{LDR analysis and results}
\subsection{Line identification}
Within the visible region of cool star spectra, there are several pairs of lines suitable for temperature determination. 
Most works are based on line pairs in the spectral domain around 6200~\AA\,(Gray \& ~Johanson 1991, Gray \& ~Brown 2001, Catalano 
et al. 2002a, 2002b, Biazzo 2006) and 6400 \AA~(Strassmeier \& Fekel ~1990, Strassmeier \& ~Schordan 2000).

\begin{figure*}
   \centering
{\includegraphics[width=15cm]{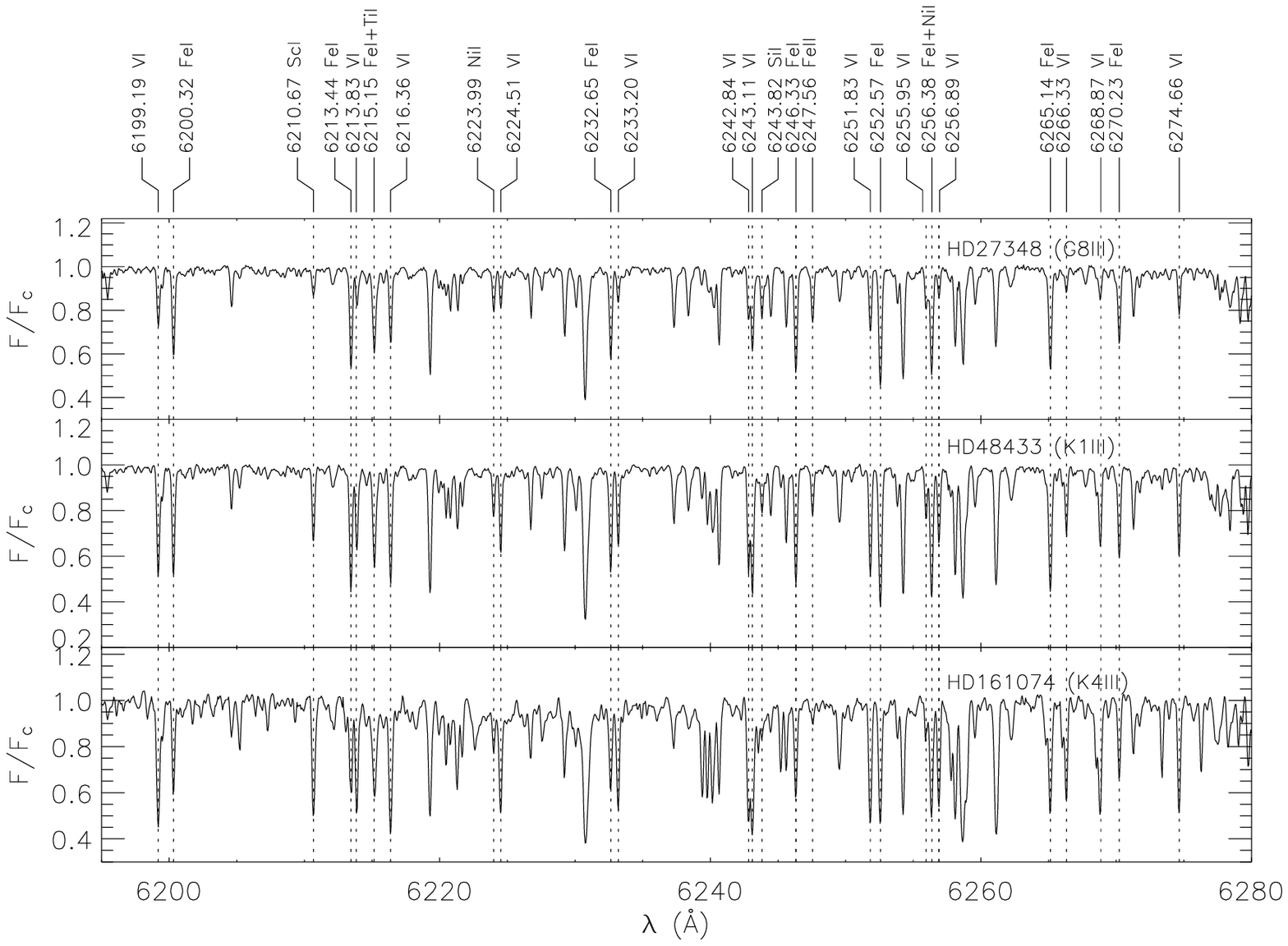}
\includegraphics[width=15cm]{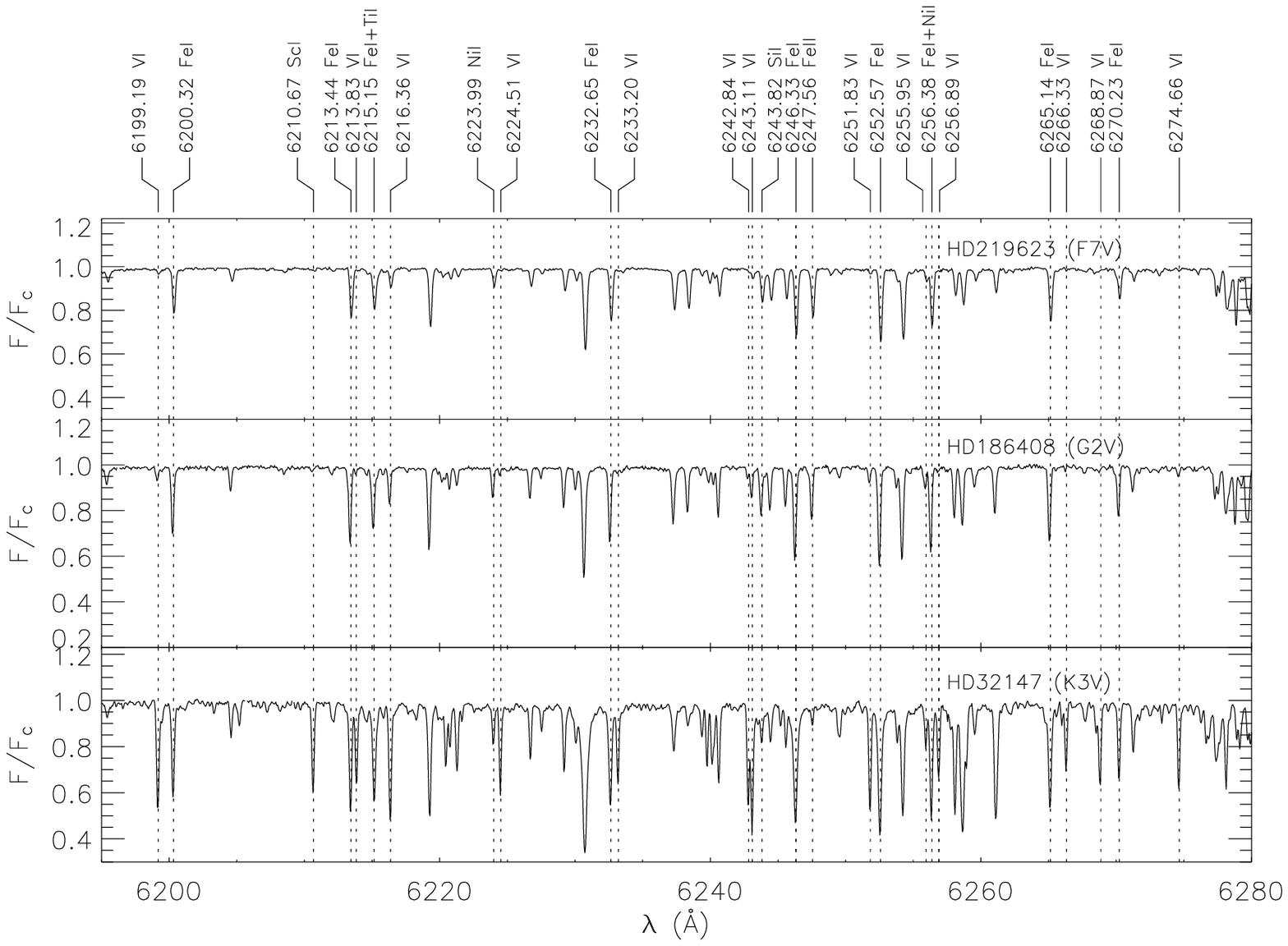}}
\caption{Three representative normalized spectra for giant (upper panel) and MS (bottom panel) stars. Temperature 
decreases from top to bottom. The spectral lines used for the LDR computation are labeled with 
their wavelength (in \AA), 
chemical element and ionization level. The lines, with low excitation potential, such as $\lambda$6210.67 \ion{Sc}{i} or 
$\lambda$6266.33 \ion{V}{i}, increase their depth with decreasing temperature, while the $\lambda$6247.56 \ion{Fe}{ii} line, 
with high excitation potential, shows an opposite behaviour. The lines with intermediate $\chi$, such as 
$\lambda$6256.35 \ion{Ni}{i}, don't display any relevant variation in this temperature range.}
\label{fig:confrontospettri_giganti_MS}
\end{figure*}

For this work, several spectral lines have been chosen in the 6190--6280~\AA\,range. 
A sample of ELODIE spectra of different spectral types is shown in Fig.~\ref{fig:confrontospettri_giganti_MS}. 
The lines were identified through the solar (Moore et al. 1966) and Arcturus (Griffin 1968) atlases, choosing those 
appearing unblended at the very high resolution of the aforementioned atlases. The chosen lines, their excitation 
potential $\chi$ (Moore et al. 1966) and ionization energy $I$ (Allen 1973) in electron volts are listed in 
Table~\ref{tab:spectral_line} together with a code number in the fifth column. The lines for each ratio are chosen to be close together 
in order to minimize the errors in the continuum setting. Fifteen line ratios are used altogether in this work. 
Each spectral line has an intensity depending specifically on 
temperature and gravity (or electronic pressure) through the line and continuum absorption coefficients ($l_{\nu}$ and 
$\kappa_{\nu}$). These parameters are related to the excitation and ionization potentials, $\chi$ and $I$, by means of  
Boltzmann and Saha equations. For these reasons, for example, the lines $\lambda$6243.11 \ion{V}{i} and $\lambda$6247.56 \ion{Fe}{ii} 
change in opposite directions with temperature, as displayed in Fig.~\ref{fig:confrontospettri_giganti_MS}, where the effect of 
temperature and gravity on line intensities is apparent.

\scriptsize
\begin{table}  
\caption{Spectral lines. Each line printed in italics is fully blended with the preceding one.}
\label{tab:spectral_line}
\begin{tabular}{ccccc}
\hline
{$\lambda$ }&{Element }&{ $\chi$ }&{ $I$ }&{ Code }\\       
 (\AA)     &          & (eV)    & (eV)    & \\\hline     
6199.19 &~~~\ion{V}{i}~~& 0.29& 6.74 & 1\\
6200.32 &~~\ion{Fe}{i}~~& 2.61& 7.87 & 2\\
6210.67 &~~\ion{Sc}{i}~~& 0.00& 6.54 & 3\\
6213.44 &~~\ion{Fe}{i}~~& 2.22& 7.87 & 4\\
6213.83 &~~~\ion{V}{i}~~& 0.30& 6.74 & 5\\
6215.15 &~~\ion{Fe}{i}~~& 4.19& 7.87 & 6\\
{\it 6215.22} &~~\ion{{\it Ti}}{i}~~& {\it 2.69}& {\it 6.82} & \\
6216.36 &~~~\ion{V}{i}~~& 0.28& 6.74 & 7\\
6223.99 &~~\ion{Ni}{i}~~& 4.10& 7.63 & 8\\
6224.51 &~~~\ion{V}{i}~~& 0.29& 6.74 & 9\\
6232.65 &~~\ion{Fe}{i}~~& 3.65& 7.87 & 10\\
6233.20 &~~~\ion{V}{i}~~& 0.28& 6.74 & 11\\
6242.84 &~~~\ion{V}{i}~~& 0.26& 6.74 & 12\\
6243.11 &~~~\ion{V}{i}~~& 0.30& 6.74 & 13\\
6243.82 &~~\ion{Si}{i}~~& 5.61& 8.15 & 14\\
6246.33 &~~\ion{Fe}{i}~~& 3.60& 7.87 & 15\\
6247.56 &~\ion{Fe}{ii}~~& 3.89&16.18 ~~& 16\\
6251.83 &~~~\ion{V}{i}~~& 0.29& 6.74 & 17\\
6252.57 &~~\ion{Fe}{i}~~& 2.40& 7.87 & 18\\
6255.95 &~~\ion{Fe}{i}~~&    ?& 7.87 & 19\\
6256.35 &~~\ion{Ni}{i}~~& 1.68& 7.63 & 20\\
{\it 6256.36} &~~\ion{{\it Fe}}{i}~~& {\it 2.45}& {\it 7.87} & \\
6256.89 &~~~\ion{V}{i}~~& 0.28& 6.74 & 21\\
6265.14 &~~\ion{Fe}{i}~~& 2.18& 7.87 & 22\\
6266.33 &~~~\ion{V}{i}~~& 0.28& 6.74 & 23\\
6268.87 &~~~\ion{V}{i}~~& 0.30& 6.74 & 24\\
6270.23 &~~\ion{Fe}{i}~~& 2.86& 7.87 & 25\\
6274.66 &~~~\ion{V}{i}~~& 0.27& 6.74 & 26\\
\hline 
\end{tabular}				      
\end{table}
\normalsize

The measure of line depths ($d$), LDRs ($r$), and the evaluation of errors has been carried out according to 
the guidelines of Catalano et al. (2002a).

\subsection{Metallicity effect}
\label{sec:metallicity}

Before proceeding with the calibration, it is important to evaluate the dependence of line-depth ratios 
on metallicity. Gray (1994) 
has investigated the influence of metallicity on color indices, finding an empirical relation between $B-V$ and the logarithm of 
the iron abundance normalized to the Sun ([Fe/H]). A calibration relation between  $B-V$ and $T_{\rm eff}$ depending on [Fe/H] 
has been also proposed by Alonso et al. (1999). However, the $B-V$ color index is only very slightly dependent on [Fe/H] for stars 
of nearly solar metallicity. Furthermore, Gray (1994) did not find any dependence on [Fe/H] for LDRs of weak lines and only a 
very weak dependence for the line pair $\lambda$6251.83 \ion{V}{i}--$\lambda$6252.57 \ion{Fe}{i}, amounting to about $\pm 20$~K 
for [Fe/H] in the range $\pm\,$0.3. Since all the stars analysed in this work have metallicities close to the solar value 
(within $\pm$\,0.3) and taking also into account the large uncertainty on many [Fe/H] literature data, no correction for this 
parameter has been applied. Anyway, given the high number of stars in our sample, possible residual effects due to small metallicity 
differences not properly accounted for are not expected to significantly affect our calibration.

\subsection{Gravity effect}

Line depth ratios are sensitive to gravity due to the dependence on the electron pressure of the H$^{-}$ bound-free 
continuum absorption coefficient. As a consequence, to determine the temperature of stars with different gravity,
such as giants and MS, it is necessary to evaluate this dependence and/or to set appropriate temperature 
scales (Biazzo 2001).

\begin{figure}[h]
  \begin{center}
  \includegraphics[width=8cm]{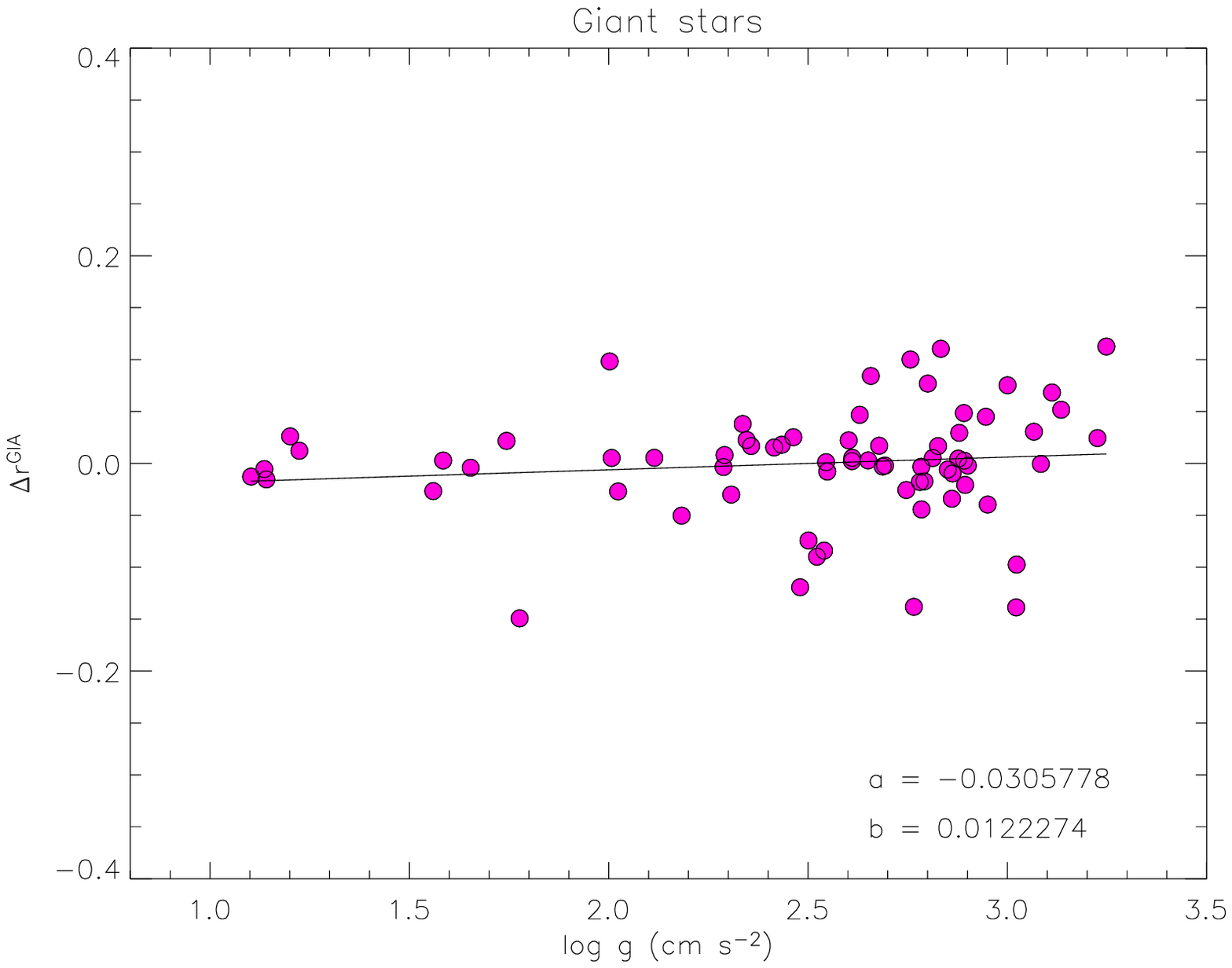}
  \includegraphics[width=8cm]{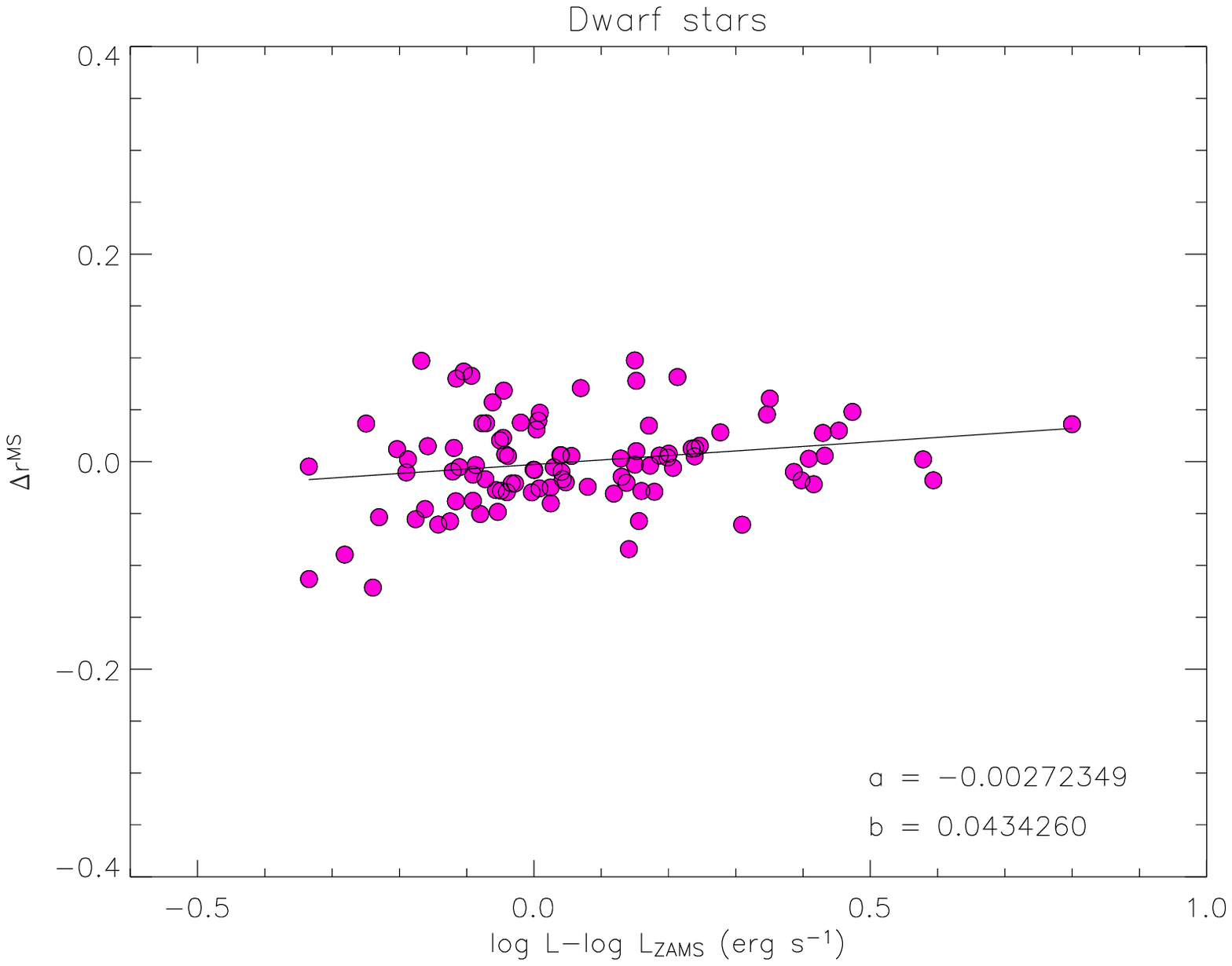}
  \end{center}
\caption{Examples of the residuals of the LDR $\lambda6252$ \ion{V}{i}-$\lambda6253$ \ion{Fe}{i} compared 
to the polynomial fit as a function of the gravity index (dots). The continuous lines represent linear fits 
to the data of giants and MS stars, respectively.}
\label{fig:gravity_dependence}
\end{figure}

\begin{figure*}
  \begin{center}
  \includegraphics[width=8.5cm]{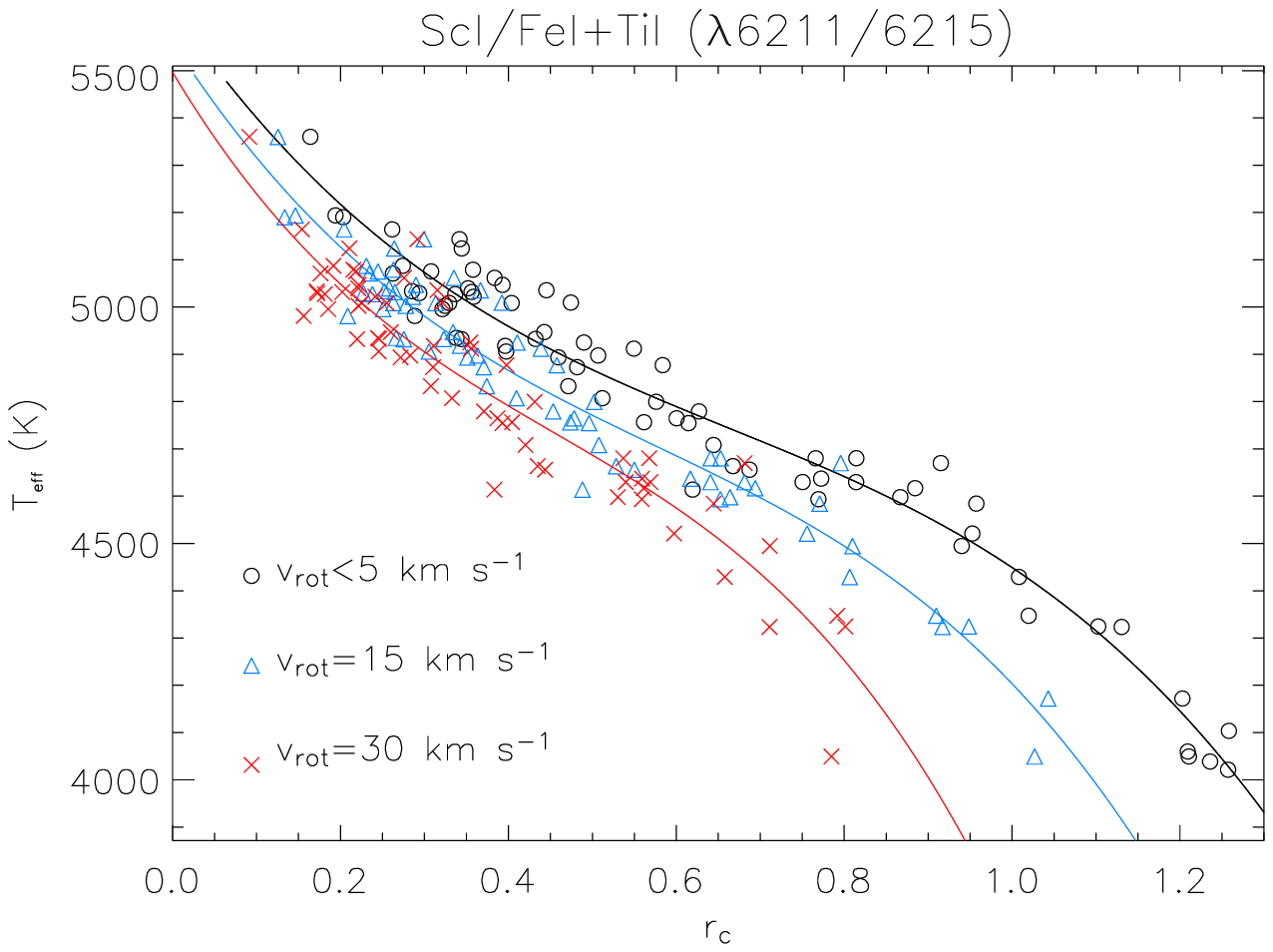}
  \includegraphics[width=8.5cm]{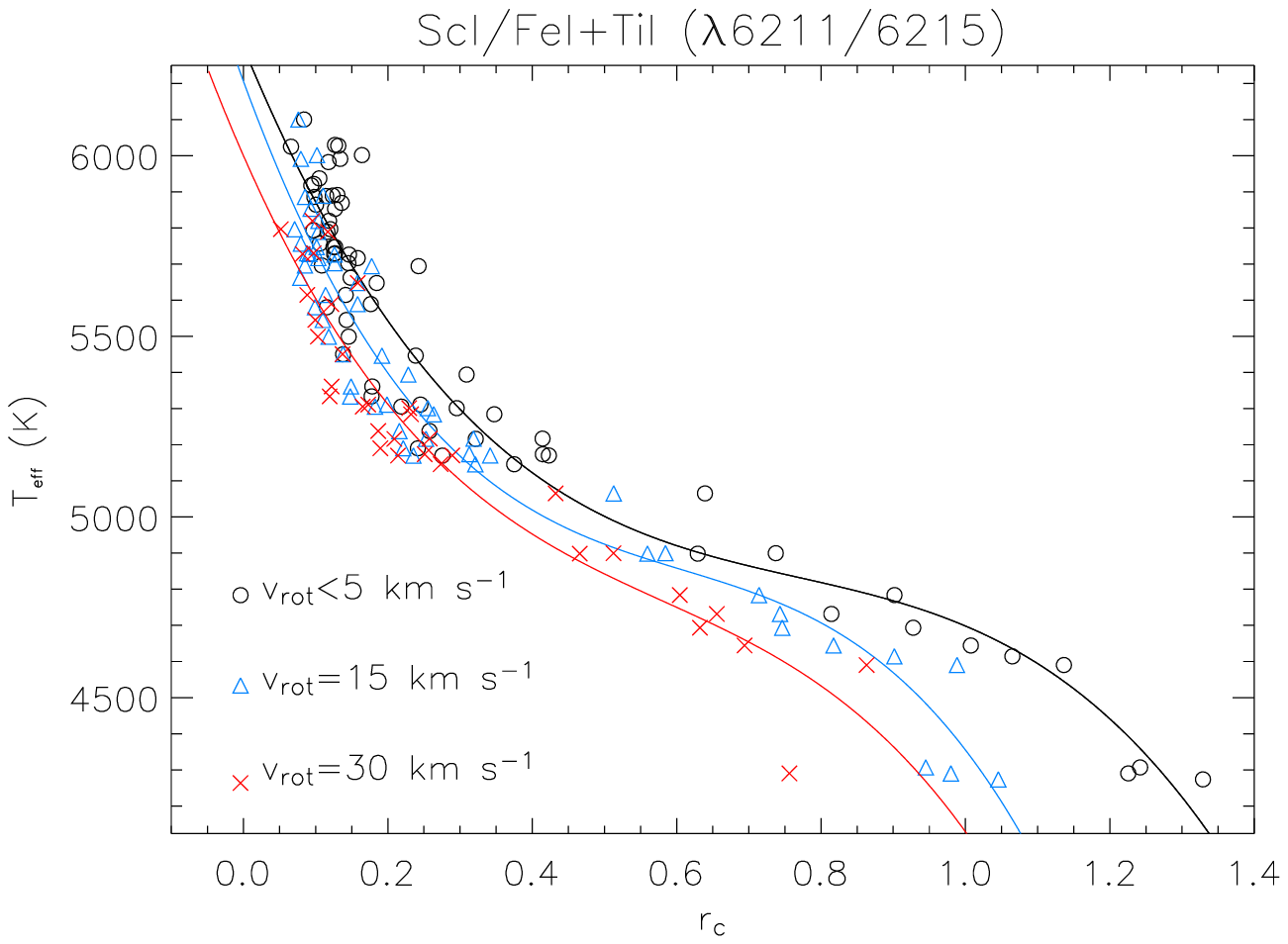}
  \end{center}
\caption{$r_{\rm c}-T_{\rm eff}$ calibration for the line pair $\lambda6211$ \ion{Sc}{i}-$\lambda6215$ 
\ion{Fe}{i}+\ion{Ti}{i} at different $v\sin i$'s (open circles: $v\sin i$=0 km s$^{-1}$; triangles: 
$v\sin i$=15 km s$^{-1}$; crosses: $v\sin i$=30 km s$^{-1}$) for giant stars ({\it left panel}) and MS 
stars ({\it right panel}).}
\label{fig:cal_ELO_MS_GIA}
\end{figure*}

For this reason, in the present work different $r-T_{\rm eff}$ calibrations have been performed for MS and giant 
stars. We have taken into account the gravity spread in each sample in the following way:
\begin{itemize}
\item[i.] for MS stars we have considered the difference $\Delta L$ between their luminosity and that of ZAMS stars
of the same color (Fig.~\ref{fig:hr_diagram_elo_red}) as a gravity index;
\item[ii.] for giant stars we have computed their gravity, $\log g$, according to the evolutionary tracks of Girardi 
et al. (2000).
\end{itemize}
Then, the residuals $\Delta r_{\rm MS, GIA}$ of LDRs compared to the fits in the $(B-V)_0-r$ plane have been plotted 
against $\Delta L$ and $\log g$ for MS and giant stars, respectively. A linear fit to the data provides very small 
positive slopes, indicating slightly increasing LDRs with gravity (Fig.~\ref{fig:gravity_dependence}). 
The linear fit was obtained by means of POLY\_FIT, an IDL routine using matrix inversion, which performs a 
weighted least-square polynomial fit with error estimates. If $a$ and $b$ are the intercept and the slope of these 
linear fits, respectively, the gravity-corrected line-depth ratio, $r_{\rm c}$, is 
\begin{eqnarray}
\label{eq:MS}
r_{\rm c}=r - (a+b\Delta L),
\end{eqnarray} 
{\noindent for MS stars and} 
\begin{eqnarray}
\label{eq:GIA}
r_{\rm c}=r - (a+b\log g), 
\end{eqnarray} 
{\noindent for giant stars. In Eq.~\ref{eq:MS} and \ref{eq:GIA}, $r$ is the observed LDR.} 
The final polynomial fits to the corrected LDRs are of the type 
\begin{eqnarray}
\label{eq:coeff}
T_{\rm eff}=c_0+c_1 r_{\rm c}+...+c_n r^n_{\rm c}, \nonumber
\end{eqnarray} 
{\noindent both for MS and giant stars, with $c_0$, ..., $c_n$ coefficients of the polynomial fits.}
An example of calibration is illustrated in Fig.~\ref{fig:cal_ELO_MS_GIA} and the polynomial fit coefficients are listed in 
Table \ref{tab:coefficients_elodie} for the 15 different line pairs.

\begin{figure}[h]
  \begin{center}
  \includegraphics[width=8cm]{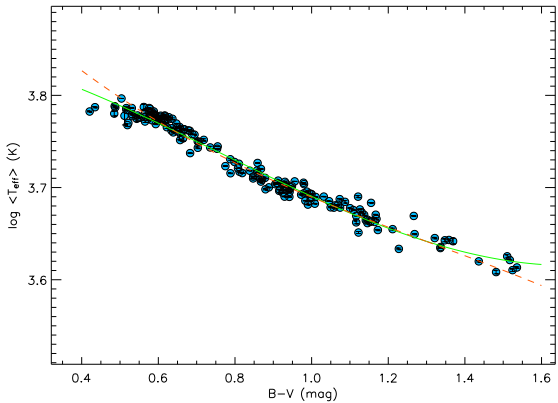}
  \end{center}
\caption{$<T_{\rm eff}>$ of the standards as a function of the de-reddened $B-V$. The continuous and dashed lines 
are referred to our and Gray's calibrations, respectively.}
\label{fig:Tm_BV}
\end{figure}

With the aim of verifying the internal consistency of our method, we have computed the effective temperature of 
each star coming from each $r_{\rm c}-T_{\rm eff}$ calibration and we have carried out the weighted average on 
all the 15 LDR. The resulting $<T_{\rm eff}>$ are plotted as a function of $(B-V)_0$ in Fig.~\ref{fig:Tm_BV}, 
together with the following polynomial fit on the points (continuous line):
\begin{eqnarray}
\log <T_{\rm eff}> & = & 3.85478-0.0582469(B-V)_0 \nonumber \\
		 & - & 0.1883(B-V)_0^2 + 0.0823(B-V)_0^3 \nonumber
\end{eqnarray}
{\noindent In the range 0.5$\ltsim (B-V)_0 \ltsim$1.5, our $<T_{\rm eff}>-(B-V)_0$ calibration is 
perfectly consistent with the one expressed in Eq.~\ref{eq:Gray_cal} and obtained by Gray 
(2005). The data scatter is obviously larger than the individual errors in $<T_{\rm eff}>$ (originating 
from the LDR errors), due to the errors in the $B-V$ measurements and to the residual dependence of $B-V$ on 
the stellar parameters other than the effective temperature, like metallicity and microturbulence.}
 
The values of $<T_{\rm eff}>$, $L$ and $r_{\rm c}$ for all the standard stars are listed in Table~5\footnote{Table~5 
is only available in electronic form at the Web site http://cdsads.u-strasbg.fr/.}. 

\subsection{Rotational broadening effect}

In a moderately rotating star, the depth at the line center decreases with increasing $v\sin i$ at a rate depending on the line 
characteristics. Following Stift \& Strassmeier (1995), in a weak line for which the saturation effects are very small, the residual 
intensity directly reflects the run of the opacity profile with frequency. A strong line, in contrast, behaves differently, 
because saturation leads to a contribution to the intensity at the line center that remains fairly high over a considerable 
part of the stellar disk, resulting in a slow decrease in the central line intensity with $v\sin i$. This implies that if 
two lines are of comparable strength and do not differ radically in the broadening parameters, the LDR will not 
depend on rotation. Stift \& Strassmeier (1995) investigated the dependence of the LDR on rotation by synthesizing the 
spectral region containing the $\lambda$6252 \ion{V}{i} and $\lambda$6253 \ion{Fe}{i} lines in a grid of atmospheric models 
(Kurucz 1993) for main-sequence stars with $v\sin i$ values ranging from 0.0 to 6.0 km s$^{-1}$. They estimated that, for a 
fixed microturbulence, the rotation dependence is always present at all temperatures (from 3500 to 6000 K)  
for a $v\sin i \ge4-5$ km s$^{-1}$. 

\begin{figure*}
  \begin{center}
  \includegraphics[width=17cm]{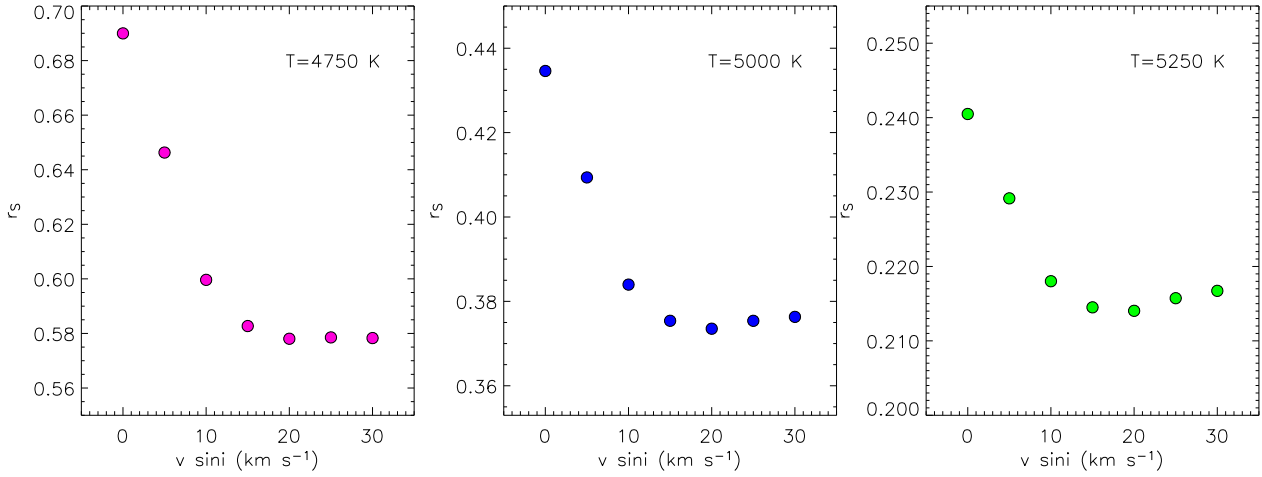}
  \end{center}
\caption{Synthetic LDRs at 3 different $T_{\rm eff}$ as a function of $v\sin i$ 
for the line pair $\lambda6211$ \ion{Sc}{i}-$\lambda6215$ \ion{Fe}{i}+\ion{Ti}{i}. The synthetic spectra 
are referred to a star with $\log g$=2.5, [Fe/H]=0.0, [$\alpha$/Fe]=0.0, $v_t$=1.8, and have been 
degraded to the same resolution of ELODIE.}
\label{fig:ldr_vsini}
\end{figure*}

We find similar results computing the LDRs obtained from high-resolution synthetic spectra. In~ 
particular, we have considered the Synthetic Stellar Library described and made available by Coelho et 
al. (2005), which ~contains spectra synthesized by adopting the model atmospheres 
of Castelli \& Kurucz (2003). These spectra are sampled at 0.02 \AA, range from 
the near-ultraviolet (300 nm) to the near-infrared (1.8 $\mu$m), and cover the following grid of 
parameters: 3500$~\le T_{\rm eff}~\le$7000 K, 0.0 $~\le\log g~\le$ 5.0, $-$2.5$\le$[Fe/H]$\le$+0.5, 
$\alpha$-enhanced [$\alpha$/Fe]=0.0, 0.4 and microturbulent velocity $v_t$=1.0, 1.8, 2.5 km s$^{-1}$. 
In Fig.~\ref{fig:ldr_vsini} we show, as an example, the behaviour of synthetic LDRs with $v\sin i$ 
at three different $T_{\rm eff}$ for the line pair $\lambda6211$ \ion{Sc}{i}-$\lambda6215$ \ion{Fe}{i}+\ion{Ti}{i}. 
The synthetic-LDR for ~this couple decreases with the increase in $v\sin i$ from 0 to about 20 km s$^{-1}$ 
and then remains nearly constant. This behaviour is apparent in all the three temperatures displayed in 
Fig.~\ref{fig:ldr_vsini}, and is more evident in the lowest one. Neglecting this effect, a systematic error 
in the effective temperature can arise. For instance, if we use the $r_{\rm c}-T_{\rm eff}$ calibration obtained 
by us at $v\sin i$=0 km s$^{-1}$ for a giant star with $T_{\rm eff}$=4750 K and $v\sin i$= 20 km s$^{-1}$, 
we overestimate its effective temperature by about 80 K. As a consequence, this LDR appears to be 
quite sensitive to the rotation velocity in the range 0--20 km s$^{-1}$. Thus, the rotational broadening 
must be taken into account to perform a reliable calibration for this as well as for several other LDRs.

To study in detail the influence of rotational broadening on LDRs, we have computed the LDRs of our star sample broadening 
the corresponding spectra at different rotation velocities from 0 to 30 km\,s$^{-1}$ in steps of 5 km\,s$^{-1}$. The LDR calculation 
has been carried out with a simple code written in IDL\footnote{IDL (Interactive Data Language) is a trademark of Research Systems 
Incorporated (RSI).}. The code first convolves the spectrum with 
the required rotational profile and then, automatically, computes the 15 LDRs in the selected wavelength range.

An example of an $r_{\rm c}-T_{\rm eff}$ calibration ($\lambda6211$ \ion{Sc}{i}-$\lambda6215$ \ion{Fe}{i}+\ion{Ti}{i}) both for giant 
and MS stars is displayed in Fig.~\ref{fig:cal_ELO_MS_GIA}, where only the calibrations for three $v\sin i$ values 
(0, 15, and 30 km s$^{-1}$) are shown. We find that for some LDRs the effects of the broadening are already visible at 
5 km s$^{-1}$, while some other LDRs are practically not affected by the rotational broadening. 
The main characteristics of these calibrations can be summarized as follows.
\begin{itemize}
\item Because of blending problems, for some pairs it is not possible to broaden the lines at any 
rotational velocity (e.g., 
$\lambda6214$ \ion{V}{i}-$\lambda6213$ \ion{Fe}{i}, $\lambda6233$ \ion{V}{i}-$\lambda6232$ \ion{Fe}{i}, 
$\lambda6242$ \ion{V}{i}-$\lambda6244$ \ion{Si}{i}, $\lambda6252$ \ion{V}{i}-$\lambda6253$ \ion{Fe}{i}, 
$\lambda6257$ \ion{V}{i}-$\lambda6255$ \ion{Fe}{i}, $\lambda6257$ \ion{V}{i}-$\lambda6256$ \ion{Fe}{i}+\ion{Ni}{i}).
\item Some LDRs seem not to suffer from rotational broadening effects in almost all the effective temperature ranges 
(for example $\lambda6199$ \ion{V}{i}-$\lambda6200$ \ion{Fe}{i}, $\lambda6214$ \ion{V}{i}-$\lambda6213$ \ion{Fe}{i}, 
$\lambda6275$ \ion{V}{i}-$\lambda6270$ \ion{Fe}{i}), as expected for weak lines.
\item Some line pairs show variation due to rotational broadening for a temperature lower than about 4500\,K and 5000\,K 
for giant and MS stars, respectively (e.g., $\lambda6243$ \ion{V}{i}-$\lambda6247$ \ion{Fe}{ii}, 
$\lambda6269$ \ion{V}{i}-$\lambda6270$ \ion{Fe}{i}).
\item Other couples display changes at any temperature (e.g., $\lambda6211$ \ion{Sc}{i}-$\lambda6215$ \ion{Fe}{i}+\ion{Ti}{i}, 
$\lambda6216$ \ion{V}{i}-$\lambda6215$ \ion{Fe}{i}+\ion{Ti}{i}, $\lambda6252$ \ion{V}{i}-$\lambda6253$ \ion{Fe}{i}, 
$\lambda6266$ \ion{V}{i}-$\lambda6265$ \ion{Fe}{i}).
\item The slope of $r_{\rm c}-T_{\rm eff}$ calibration in some ratios decreases with the increase in the rotation 
(e.g., $\lambda6243$ \ion{V}{i}-$\lambda6246$ \ion{Fe}{i}), while in some other ratios increases with the rotation 
(e.g., $\lambda6216$ \ion{V}{i}-$\lambda6215$ \ion{Fe}{i}+\ion{Ti}{i}).
\item In any case, the variation of the slope with $v\sin i$ seems to be more evident for giant stars compared to 
MS stars (e.g., $\lambda6243$ \ion{V}{i}-$\lambda6246$ \ion{Fe}{i} and 
$\lambda6269$ \ion{V}{i}-$\lambda6270$ \ion{Fe}{i}). This behaviour can be due to the fact that the giants 
have narrower lines compared to MS stars, due to their lower atmospheric density. As a consequence, 
they appear to be sensitive to the rotational broadening, while the gravity broadening in MS stars is comparable 
to the rotational broadening at low rotational velocity and the effect on LDR comes out to be less prominent.
\end{itemize}

\subsubsection{Temperature sensitivity on LDR}
In order to measure the temperature sensitivity of each LDR, the slopes $\frac{{\rm d}T}{{\rm d}r_{\rm c}}$ of the polynomial 
fits have been calculated. The $\frac{{\rm d}T}{{\rm d}r_{\rm c}}$ absolute values at a temperature of 5000~K are listed 
in Table \ref{tab:coefficients_elodie} for the calibrations of the giant and MS stars not rotationally 
broadened. Values of about 10--30\,K and, in some case, even smaller, have been found for a 0.01 variation of $r_{\rm c}$, which 
represents the typical uncertainty for the LDR determination in well-exposed spectra. Stars below 4000~K and above 6200~K 
have the most uncertain temperatures because of the influence of molecular bands in the coolest stars 
and the very small depths of the low-excitation lines in the hottest stars of our sample, respectively. 

\begin{figure}	
{\includegraphics[width=8cm]{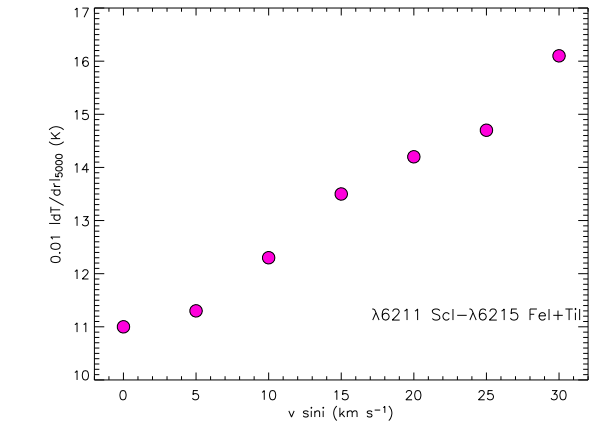}}
\caption{Example of variation of the temperature sensitivity with the rotational velocity obtained for the giant calibration 
$\lambda6211$ \ion{Sc}{i}-$\lambda6215$ \ion{Fe}{i}+\ion{Ti}{i}.}
\label{fig:dT_dr_1115}
\end{figure}

Thus, the temperature sensitivity changes from an LDR to another one, but it is also a function of the rotational velocity. 
Fig.~\ref{fig:dT_dr_1115} displays an example of the variation of the temperature sensitivity with the rotational 
velocity of the couple 
$\lambda6211$ \ion{Sc}{i}-$\lambda6215$ \ion{Fe}{i}+\ion{Ti}{i} obtained for the giant calibration. For this ratio, the temperature 
sensitivity decreases with the increasing rotational broadening. 
As we mentioned before, other LDRs display different 
behaviours of the temperature sensitivity as a function of $v\sin i$.

The coefficients ($c_0, ..., c_n$) of the fits and the values of $\frac{{\rm d}T}{{\rm d}r_{\rm c}}$ and rms obtained at 
$v\sin i$=0, 15, and 30 km s$^{-1}$ are listed in Tables~\ref{tab:coefficients_elodie},~\ref{tab:coefficients_elodie_15}
~\ref{tab:coefficients_elodie_30}, respectively.

\begin{table*} 
\caption{Coefficients of the fits $T_{\rm eff}=c_0+c_1 r_{\rm c}+...+c_n r^n_{\rm c}$ for the ELODIE spectra not 
rotationally broadened.}
\label{tab:coefficients_elodie}
\begin{center}
\begin{tabular}{crrrrrrrr}
\hline
\multicolumn{9}{c}{G{\sc iant} S{\sc tars}}\\
\hline
LDR&1/2&3/6&5/4&7/6&9/8&11/10&12/14&13/15\\
\hline
$c_0$& 5658.52& 5630.37& 5357.34& 8808.49& 4976.68& 5328.31& 5375.97& 5435.02\\
$c_1$&$-$1565.89&$-$2563.00&$-$846.737&$-$10484.1& 697.119&$-$749.330&$-$292.110&$-$8.57992\\
$c_2$& 1384.63& 2766.46&$-$176.810& 10403.9&$-$849.779&$-$150.672&$-$25.7998&$-$654.941\\
$c_3$&$-$807.719&$-$1384.76&$-$80.6700&$-$3804.58& 184.978&$-$3.71716&0.715242&      \\
$0.01\,|\frac{{\rm d}T}{{\rm d}r_{\rm c}}|_{5000\rm\,K}$ &10.4&11.0&12.4&14.9& 5.2&10.9& 4.0&12.8\\
rms&0.080&0.060&0.070&0.071&0.081&0.078&0.085&0.068\\
\hline
LDR&13/16&17/18&21/19&21/20&23/22&24/25&26/25&\\
\hline
$c_0$&   5479.94& 5630.09& 5204.23& 5407.48& 5324.25& 5346.80& 5335.75&\\
$c_1$&$-$325.954&$-$1935.08&$-$53.5851&$-$2091.73&$-$1625.05&$-$1054.34&$-$327.402&\\
$c_2$&   4.71410& 2496.75&$-$325.674& 2640.38& 1431.70& 619.856&$-$313.672&\\
$c_3$&   1.43373&$-$1976.03& 74.5653&$-$1956.14&$-$1193.13&$-$343.396&$-$42.0121&\\
$0.01\,|\frac{{\rm d}T}{{\rm d}r_{\rm c}}|_{5000\rm\,K}$ &4.1&13.2& 6.3&12.9&13.9&9.2&10.1&\\
rms&0.070&0.054&0.068&0.062&0.063&0.081&0.081\\
\hline
\multicolumn{9}{c}{D{\sc warf} S{\sc tars}}\\
\hline
LDR&1/2&3/6&5/4&7/6&9/8&11/10&12/14&13/15\\
\hline
$c_0$&   6592.19  &   6298.64& 6303.41& 8075.38& 6181.66& 6204.32& 6122.76& 6583.91\\
$c_1$&$-$4053.01&$-$4813.87&$-$5018.31&$-$6053.31&$-$1077.50&$-$3855.13&$-$895.008&$-$3240.32\\
$c_2$&   4020.50  &   5670.56& 6604.68& 4843.37& 197.649& 3379.47& 158.882& 2763.93\\
$c_3$&$-$1663.63&$-$2457.28&$-$3562.99&$-$1688.39&$-$7.73603&$-$1175.00&$-$8.65057&$-$1211.49\\
$0.01\,|\frac{{\rm d}T}{{\rm d}r_{\rm c}}|_{5000\rm\,K}$ &11.0&10.1&13.4&15.7&4.5&12.4&2.8&13.2\\
rms&0.065&0.060&0.040&0.057&0.085&0.047&0.087&0.049\\
\hline
LDR&13/16&17/18&21/19&21/20&23/22&24/25&26/25&\\
\hline
$c_0$&   6250.15&   6486.68& 6178.37&	6210.10&   6228.96&   6120.07&   6271.19&\\
$c_1$&$-$732.154&$-$5065.09&$-$1203.98&$-$5244.93&$-$6031.24&$-$2734.59&$-$1994.09&\\
$c_2$&   143.714&   7059.62& 261.964&	6959.06&   9450.38&   1909.74&   716.031&\\
$c_3$&$-$9.97940&$-$4105.26&$-$19.2823&$-$3796.17&$-$5981.44&$-$559.915&$-$138.664&\\
$0.01\,|\frac{{\rm d}T}{{\rm d}r_{\rm c}}|_{5000\rm\,K}$ &1.4&15.2& 5.0&16.5&18.0& 8.5& 9.8&\\
rms&0.074&0.043&0.067&0.047&0.053&0.067&0.076&\\
\hline
\end{tabular}				      
\end{center}
\end{table*}
\normalsize

\begin{table*}	
\caption{Coefficients of the fits $T_{\rm eff}=c_0+c_1 r_{\rm c}+...+c_n r^n_{\rm c}$ for the ELODIE spectra  
rotationally broadened to 15 km s$^{-1}$. The empty columns relate to the line pairs completely blended at this 
rotational velocity.}
\label{tab:coefficients_elodie_15}
\begin{center}
\begin{tabular}{crrrrrrrr}
\hline
\multicolumn{9}{c}{G{\sc iant} S{\sc tars}}\\
\hline
LDR&1/2&3/6&5/4&7/6&9/8&11/10&12/14&13/15\\
\hline
$c_0$& 5249.84& 5559.50& & 8090.67& 5453.71& 5355.55& & 5662.56\\
$c_1$& 245.810&$-$2741.36& &$-$9105.81&$-$489.599&$-$879.618& &$-$670.393\\
$c_2$&$-$1064.84& 3273.61& & 9705.52&$-$34.6856&$-$138.541& &$-$13.6821\\
$c_3$& 306.601&$-$1888.48& &$-$3948.66& 33.7288& 35.1345& &	   \\
$0.01\,|\frac{{\rm d}T}{{\rm d}r_{\rm c}}|_{5000\rm\,K}$ &9.8&13.5&  &15.6&5.1&12.6&  & 8.0\\
rms&0.090&0.057&&0.065&0.087&0.068& &0.070\\
\hline
LDR&13/16&17/18&21/19&21/20&23/22&24/25&26/25&\\
\hline
$c_0$&    5400.03&    5051.44& & 5605.92& 5297.24& 5237.87& 5202.02&\\
$c_1$& $-$231.533&$-$1505.89& &$-$3632.92&$-$1585.45&$-$526.586& 203.540&\\
$c_2$&    13.1768&   4285.89& & 5882.55& 940.850&$-$111.045&$-$1183.49&\\
$c_3$&$-$0.368023&$-$1829.64& &$-$4511.96&$-$817.658& 47.8455& 391.671&\\
$0.01\,|\frac{{\rm d}T}{{\rm d}r_{\rm c}}|_{5000\rm\,K}$ &2.4&15.5&  &15.9&15.8& 7.9&10.0&\\
rms&0.072&0.055& &0.050&0.056&0.087&0.088&\\
\hline
\multicolumn{9}{c}{D{\sc warf} S{\sc tars}}\\
\hline
LDR&1/2&3/6&5/4&7/6&9/8&11/10&12/14&13/15\\
\hline
$c_0$& 6537.09& 6205.81& & 7824.72& 6216.99& 6451.17& & 6522.90\\
$c_1$&$-$3721.86&$-$5454.44& &$-$5415.18&$-$1442.34&$-$5424.86& &$-$2833.56\\
$c_2$& 3167.41& 7964.98& & 3678.15& 532.435& 5770.74& & 1687.54\\
$c_3$&$-$1119.43&$-$4361.33& &$-$1198.04&$-$70.4289&$-$2439.22& &$-$485.220\\
$0.01\,|\frac{{\rm d}T}{{\rm d}r_{\rm c}}|_{5000\rm\,K}$ &10.7&11.8&  &17.9&3.2&16.0&  &11.0\\
rms&0.061&0.054&&0.048&0.089&0.036& &0.053\\
\hline
LDR&13/16&17/18&21/19&21/20&23/22&24/25&26/25&\\
\hline
$c_0$&   6194.16&   6312.44& & 6347.22& 5964.93& 5988.72& 6264.03&\\
$c_1$&$-$549.734&$-$4404.66& &$-$7231.99&$-$4180.21&$-$2177.96&$-$2135.85&\\
$c_2$&   83.9286&   5255.34& & 12331.5& 4047.15& 1527.77& 1057.32&\\
$c_3$&$-$4.27056&$-$3379.88& &$-$8240.08&$-$1716.93&$-$483.352&$-$322.180&\\
$0.01\,|\frac{{\rm d}T}{{\rm d}r_{\rm c}}|_{5000\rm\,K}$ &1.0&18.6&  &16.4&18.7& 6.9&10.1&\\
rms&0.075&0.033& &0.053&0.049&0.082&0.087&\\
\hline
\end{tabular}				      
\end{center}
\end{table*}
\normalsize

\begin{table*} 
\caption{Coefficients of the fits $T_{\rm eff}=c_0+c_1 r_{\rm c}+...+c_n r^n_{\rm c}$ for the ELODIE spectra 
rotationally broadened to 30 km s$^{-1}$. The empty columns relate to the line pairs completely blended at this 
rotational velocity.}
\label{tab:coefficients_elodie_30}
\begin{center}
\begin{tabular}{crrrrrrrr}
\hline
\multicolumn{9}{c}{G{\sc iant} S{\sc tars}}\\
\hline
LDR&1/2&3/6&5/4&7/6&9/8&11/10&12/14&13/15\\
\hline
$c_0$& 5481.25& 5496.39& & 7025.27& & & & 5695.10\\
$c_1$&$-$596.797&$-$2961.51& &$-$5597.99& & & &$-$800.163\\
$c_2$&$-$43.7139& 4228.71& & 5957.99& & & & 75.3450\\
$c_3$&$-$48.3037&$-$3087.51& &$-$2710.79& & & &  \\
$0.01\,|\frac{{\rm d}T}{{\rm d}r_{\rm c}}|_{5000\rm\,K}$ &9.3&16.1&  &16.5&  &  &  & 8.4\\
rms&0.095&0.058&&0.061&&& &0.075\\
\hline
LDR&13/16&17/18&21/19&21/20&23/22&24/25&26/25\\
\hline
$c_0$&    5337.08& & &  & 5150.51&   5190.59&   5357.17&\\
$c_1$& $-$166.813& & & &$-$825.095&$-$500.186&$-$591.566&\\
$c_2$&    8.35369& & & &  70.5176&28.5439&$-$202.882&\\
$c_3$&$-$0.153085& & & &$-$989.250& 6.34586&   92.0652&\\
$0.01\,|\frac{{\rm d}T}{{\rm d}r_{\rm c}}|_{5000\rm\,K}$ &1.7&  &  &  &15.1& 7.5& 9.5&\\
rms&0.075& & & &0.071&0.089&0.085&\\
\hline
\multicolumn{9}{c}{D{\sc warf} S{\sc tars}}\\
\hline
LDR&1/2&3/6&5/4&7/6&9/8&11/10&12/14&13/15\\
\hline
$c_0$&   6432.25&   5999.97& &   7479.52& & & & 6874.46\\
$c_1$&$-$3039.71&$-$4545.30& &$-$3706.65& & & &$-$3697.12\\
$c_2$&   2161.86&   6247.09& &   1147.46& & & & 2312.46\\
$c_3$&$-$692.710&$-$3573.28& &$-$116.692& & & &$-$595.265\\
$0.01\,|\frac{{\rm d}T}{{\rm d}r_{\rm c}}|_{5000\rm\,K}$ &10.3&11.6&  &18.9&  &  &  &11.1\\
rms&0.066&0.046&&0.051&&& &0.079\\
\hline
LDR&13/16&17/18&21/19&21/20&23/22&24/25&26/25&\\
\hline
$c_0$&    6082.76& & & &   5895.59&   5712.75&   6039.12&\\
$c_1$& $-$329.412& & & &$-$4065.13&$-$1362.68&$-$885.351&\\
$c_2$&    30.3591& & & &  5641.78&   931.915&$-$863.622&\\
$c_3$&$-$0.945513& & & &$-$5297.51&$-$312.388&   527.239&\\
$0.01\,|\frac{{\rm d}T}{{\rm d}r_{\rm c}}|_{5000\rm\,K}$ &1.0&   &  &  &20.0& 5.0& 10.1&\\
rms&0.080&& & &0.033&0.085&0.090&\\
\hline
\end{tabular}				      
\end{center}
\end{table*}
\normalsize

\section{Conclusion}
In this work we have performed LDR--$T_{\rm eff}$ calibrations from high-resolution spectra taking into account 
the corrections for the gravity effect and the rotational broadening. To our knowledge, this is the first work in which 
the dependence of LDRs on rotational velocity has been considered on real spectra. In previous studies, this dependence had 
not been taken ~into account, also~ following ~Gray \& Johanson's (1991) early suggestions that considered the 
effect of $v\sin i$ to be negligible. Here we demonstrate that the rotational broadening effect is already evident 
at 5 km s$^{-1}$ both in synthetic and real spectra of cool stars. Its effect can be neglected only in particular 
temperature domains and for a few line pairs. In other situations, the dependency of LDRs on $v\sin i$ must be properly 
taken into account, at least at a spectral resolution as high as 42\,000. We show that neglecting this effect can lead 
to temperature overestimates as high as $\approx$100 K. 

This paper provides calibrations that can be used for temperature determinations of stars with 0$\le v\sin i\ltsim$30 
km s$^{-1}$ and observed with ELODIE as well as with other spectrographs at a similar resolution. We have shown that all 
the LDRs examined in the present paper display a good sensitivity to the effective temperature which allows us to reach 
an uncertainty lower than 10--15 K at $R=42\,000$ for stars rotating up to $v\sin i$=30 km s$^{-1}$. The simultaneous 
use of several LDRs allows us to improve the temperature sensitivity and to fully detect and analyse effective temperature 
modulations produced by cool starspots in active stars, as we have already shown in previous works (Frasca et al. 2005, 
Biazzo et al. 2007).

\acknowledgements
This work has been supported by the Italian National Institute for Astrophysics (INAF), by the Italian 
{\em Ministero dell'Istruzione, Universit\`a e  Ricerca} (MIUR) and by the 
{\em Regione Sicilia} which are gratefully acknowledged. We thank the referee for useful suggestions.
We are also grateful to  Mrs. Luigia Santagati for the English revision of the text.
This research has made use of SIMBAD and VIZIER databases, 
operated at CDS, Strasbourg, France.

\end{document}